\documentclass[ejsv2,preprint,noshowframe]{imsart}
\RequirePackage[numbers]{natbib}
\RequirePackage{amsmath,graphicx,booktabs,longtable,array,calc}
\RequirePackage[colorlinks,citecolor=blue,urlcolor=blue]{hyperref}
\providecommand{\tightlist}{\setlength{\itemsep}{0pt}\setlength{\parskip}{0pt}}

\startlocaldefs
\setpkgattr{figurename}{size}{\upshape}
\setpkgattr{tablename}{size}{\upshape}
\newtheorem{theorem}{Theorem}
\newtheorem{proposition}{Proposition}
\newtheorem{corollary}{Corollary}
\newtheorem{remark}{Remark}
\endlocaldefs
\begin{document}
\begin{frontmatter}
\title{Functional-Specific Information Borrowing under Limited Overlap}
\runtitle{Functional-specific borrowing}
\begin{aug}
\author{\fnms{Shinichiro}~\snm{Shirota}}
\runauthor{S. Shirota}
\address{Hitotsubashi University}
\end{aug}
\begin{abstract}Source observations can improve a target average without estimating the entire shared response accurately. We study this distinction in two independent observational samples whose treatment effects share a rough component and differ by a smoother discrepancy. For power-law treatment overlap known for the analysis, known design density, fixed positive Sobolev radii, and Gaussian noise, we derive matching minimax mean squared error bounds with three regimes. Two explicit source-information thresholds determine the onset of rate improvement and attainment of the source-known rate. A signed local correction explains the thresholds: source observations estimate only the part of the integration functional poorly recovered from the target sample. Direct likelihood lower bounds apply to all estimators. Existing joint reproducing-kernel recovery, which does not require propensity inputs, attains these rates, and standard bias-aware intervals are conditionally honest under the specified class. A supplementary zero-overlap result separates estimation error from residual identification uncertainty. Numerical results illustrate both gains and losses when the total number of participants is fixed. The intermediate rate depends on overlap geometry, while smoothness determines its two endpoints.\end{abstract}
\end{frontmatter}
\section{Introduction}\label{introduction}

A source population may improve estimation of a target average even when
the populations have different treatment effects. How much source
information is needed before this improvement changes the minimax rate?
When does its benefit saturate? These questions depend on the target
functional and on where target information is scarce.

We consider a one-dimensional benchmark in which treated target
observations become rare near a boundary. The target effect is \(g+d\),
the source effect is \(g\), and the discrepancy \(d\) is smoother than
\(g\). The estimand remains the target average \(L(g+d)\). Our main
result gives two source-information thresholds and three matching
minimax MSE orders. The gap between the thresholds is positive whenever
the discrepancy has greater Sobolev smoothness. Below the first
threshold, borrowing does not improve the order; above the second, the
uncertainty in the discrepancy is limiting.

The mechanism is a localized correction. A signed target weight \(W_h\)
avoids the least informative region while reproducing low-order moments.
Source data estimate \((L-W_h)g\), whose support has probability mass of
order \(h^{\beta+1}\). Their variance contribution is therefore
\(h^{\beta+1}/M\), where \(M=m\eta\) is source information. This is
smaller than the \(1/M\) scale of estimating the complete source
average. The discrepancy contributes the residual bias. Balancing these
terms explains both thresholds.

\subsection{Relation to existing
theory}\label{relation-to-existing-theory}

Limited overlap can produce irregular identification and unstable inverse weighting
\citep{khan,crump,rothe,mawang}. Here the target population average is kept
fixed while signed moment correction controls boundary bias. Optimal recovery
\citep{donoho}, honest interval adaptation \citep{cailow}, bias-aware inference
\citep{ak18,ak21}, and minimax linear weighting \citep{hirshberg} provide the
estimation and inference framework.

Mou et al.\ \citep[Theorems 1--3 and Corollary 1]{mou} give instance-dependent
bounds for observational linear functionals. Their first-order Sobolev
example has the same target-only endpoint as ours at $s_g=1,\beta=0$.
Our fixed-quota two-sample calculation resolves the additional source-information
axis. Wang et al.\ \citep[Theorems 4.4--4.5, arXiv v3]{wang} obtain
whole-function transfer rates through pooling and target correction. Our
integral loss and treatment-dependent design require direct functional
lower bounds. The cited theorem numbers refer to the specified preprints.

Transfer classification \citep{kpotufe,caiwei} and regression under covariate
shift \citep{pathak,mapathak} characterize related distribution-dependent
benefits. Data-fusion methods can also use trials to correct observational
confounding \citep{kallus,colnet}; our two domains instead satisfy their own
unconfoundedness assumptions and may have different treatment effects.

\paragraph{Scope.} The contribution is the explicit three-regime rate and
localized explanation for the stated scalar functional and product class.
The recovery algorithm and Gaussian calibration use existing methods.
Guarantees concern fixed known radii and the specified one-dimensional
model; estimation of the target integration law and honest adaptation to
unknown radii require additional analysis. Finite-sample comparisons below
measure average conditional worst risk.

\section{Model and criteria}\label{model-and-criteria}

\subsection{Two observational
samples}\label{two-observational-samples}

All observations are recruited within a prespecified subgroup \(G=1\).
In both domains,

\[p_\beta(x)=(\beta+1)x^\beta,\quad 0\le x\le1,\quad \beta\ge0,
\qquad Lf=\int_0^1p_\beta(x)f(x)\,dx.
\tag{1}\]

There are \(n\) independent target observations and \(m\) independent
source observations, independent across samples. For \(j=T,S\), observe
\(X,A,Y\), where \(A\mid X\sim\mathrm{Bernoulli}(e_j(X))\), and

\[e_T(x)=x^\kappa/2,\quad e_S(x)=\eta(1+x),\quad 0<\eta\le1/3,
\tag{2}\] \[Y_T=b_T(X)+A\{g(X)+d(X)\}+\epsilon_T,
\qquad Y_S=b_S(X)+Ag(X)+\epsilon_S.
\tag{3}\]

Errors are independent \(N(0,1)\), independent of the designs.
Consistency and conditional unconfoundedness within each domain give the
causal interpretation

\[\theta=E_T[Y(1)-Y(0)\mid G=1]=L(g+d).
\tag{4}\]

The propensity functions describe observational assignment, not
treatment chosen by the researcher. The density, propensities, error
variance, smoothness orders, and radii are known for the analysis.
The joint recovery estimator uses the density through $L$, the spaces,
radii and noise variance, but does not take the propensities as inputs. The sample budget
counts eligible participants, not screening costs.

\subsection{Product class}\label{product-class}

For integer \(s\ge1\), use the standard, nonhomogeneous Sobolev norm

\[\|f\|_{H^s}^2=\sum_{j=0}^s\int_0^1|f^{(j)}(x)|^2\,dx.\]

With fixed \(B,R,\rho>0\), let

\[\mathcal F=\{(b_T,b_S,g,d):\|b_T\|_{H^1},\|b_S\|_{H^1}\le B,
\ \|g\|_{H^{s_g}}\le R,\ \|d\|_{H^{s_d}}\le\rho\}.
\tag{5}\]

The main theorem assumes integers \(s_d>s_g\ge1\). Radii remain fixed as
sample sizes grow. The target-only projection is the Minkowski sum
\(RB_{H^{s_g}}+\rho B_{H^{s_d}}\), not an \(H^{s_d}\) ball. Smoothness
equivalence with a fixed-scale Matérn RKHS does not imply equality of
norms or of finite-sample bias bounds. The exact computations below use
(5).

\subsection{Quantifiers}\label{quantifiers}

Let \(P_f^{n,m}\) denote the full random-design experiment and let \(D\)
contain all covariates and assignments. Define

\[\mathfrak R_{n,m}=\inf_{\widehat\theta}\sup_{f\in\mathcal F}
E_f(\widehat\theta-\theta_f)^2,
\qquad
\mathfrak L_{n,m}=\inf_{C\in\mathcal C_\alpha}\sup_{f\in\mathcal F}E_f|C|,
\tag{6}\]

where the first infimum includes nonlinear estimators and

\[\mathcal C_\alpha=\{C:\inf_{f\in\mathcal F}P_f(\theta_f\in C)\ge1-\alpha\},
\quad 0<\alpha<1/2.\]

The interval criterion is unconditional uniform coverage. The proposed
interval satisfies the stronger coverage requirement conditional on
\(D\), almost surely, for every \(f\in\mathcal F\).

For a fixed small \(c_0>0\), let
\(\mathcal U_{c_0}(0)=\{f\in\mathcal F:\mathrm{KL}(P_f^{n,m},P_0^{n,m})\le c_0\}\).
We also consider minimax MSE over this local set, and expected length at
zero among intervals honest over all of \(\mathcal F\). We make no local
claim at arbitrary boundary points of the norm balls.

For a specified estimator $\widehat\theta=v(D)^\top Y$, define its conditional
mean squared error by $R(f,D)=E_f[(\widehat\theta-\theta_f)^2\mid D]$.
For any design-dependent linear weights, the distinction

\[\sup_f E_D R(f,D)\ \le\ E_D\sup_f R(f,D)
\tag{7}\]

is essential. Our simulations estimate the right side for specified
procedures; it is not by definition an unconditional minimax risk. The
rates in the main theorem average over design. They do not hold
uniformly over all realized designs.

\section{Main result}\label{main-result}

Write

\[\begin{aligned}t&=\kappa-\beta-1,\quad u_s=2s+2\beta+1,\\ D_s&=2s+\beta+\kappa=u_s+t,
\\ a_s&=u_s/D_s,\quad q_s=(2s+\beta)/D_s,\quad M=m\eta.\end{aligned}
\tag{8}\]

\begin{theorem}[Three minimax regimes]\label{thm:main}

Fix \(\beta\ge0\), \(\kappa>\beta+1\), integers \(s_d>s_g\ge1\), and
\(B,R,\rho>0\). There exist \(0<c<C<\infty\) and \(n_0\), depending only
on these fixed quantities, such that for every integer \(n\ge n_0\),
integer \(m\ge1\), and \(0<\eta\le1/3\) with \(M=m\eta\ge1\), the
minimax MSE lies between \(c\,r(n,M)\) and \(C\,r(n,M)\). Here

\[\mathfrak R_{n,m}\asymp r(n,M):=
\begin{cases}
n^{-a_{s_g}},&M\le n^{q_{s_g}},\\
n^{-(\beta+1)/\kappa}M^{-t/\kappa},&n^{q_{s_g}}\le M\le n^{q_{s_d}},\\
n^{-a_{s_d}},&M\ge n^{q_{s_d}}.
\end{cases}
\tag{9}\]

In particular the upper regime includes \(M\ge n\). Further,
\(\mathfrak L_{n,m}\asymp\sqrt{r(n,M)}\). The interval constants may depend on $\alpha$. All constants are independent
of $n,m,\eta$ for the fixed parameters above.
\end{theorem}

\begin{remark}[Local consequence]
The MSE order also holds on $\mathcal U_{c_0}(0)$, and every globally honest
interval has $E_0|C|\gtrsim\sqrt{r(n,M)}$. This is the zero-center
consequence of the same testing alternatives (Supplement A, Appendix C),
with constants depending on $c_0$; it is not an instance-optimality result
at arbitrary functions. Uniformity across vanishing radii or at
$\kappa=\beta+1$ is outside Theorem 1.
\end{remark}

With \(g\) known, minimax MSE is \(n^{-a_{s_d}}\). With only \(n\)
target observations it is \(n^{-a_{s_g}}\). Thus source information
first changes the order at \(n^{q_{s_g}}\), and reaches the source-known
order at \(n^{q_{s_d}}\). Equality up to constant factors at a threshold
is not a strict order improvement; the latter requires divergence of the
corresponding information ratio. At the upper threshold,
\(a_{s_d}-q_{s_d}=(\beta+1)/D_{s_d}>0\).

The intermediate expression is independent of both $s_g$ and $s_d$:
smoothness determines its endpoints, while $(\beta,\kappa)$ determines its
slope. Writing $r_{\rm low},r_{\rm mid},r_{\rm high}$ for the three
expressions in (9), throughout the intermediate regime,
\begin{equation}
\frac{r_{\rm low}}{r_{\rm mid}}=
 (M/n^{q_{s_g}})^{t/\kappa},\qquad
\frac{r_{\rm mid}}{r_{\rm high}}=
 (n^{q_{s_d}}/M)^{t/\kappa}.\tag{10}
\end{equation}
These are identities between rate expressions, rather than exact finite-sample risk ratios.
\begin{figure}[ht]
\centering\includegraphics[width=.70\linewidth]{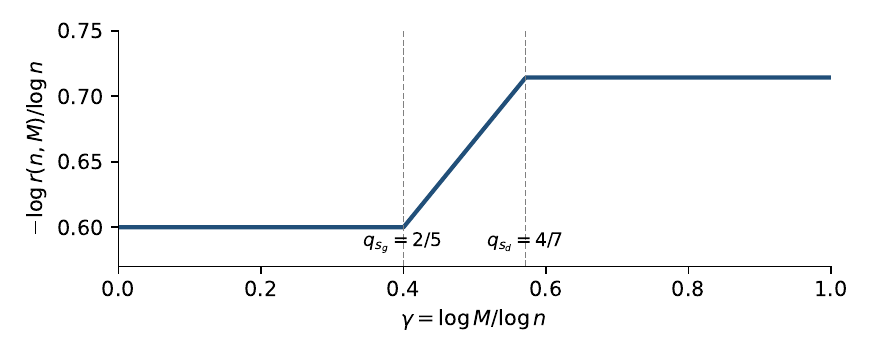}
\caption{Theoretical rate exponent for $M=n^\gamma$ at $(s_g,s_d,\beta,\kappa)=(1,2,0,3)$.
Smoothness fixes the two thresholds. This is an algebraic diagram, not an empirical fit.}
\end{figure}

\begin{corollary}[Fixed total participant budget]
Fix $0<\eta\le1/3$ and the parameters of Theorem 1. With $N=n+m$,
every allocation satisfying $n\asymp N$ and $M=m\eta\gtrsim N^{q_{s_d}}$
is rate-optimal, attaining minimax MSE of order $N^{-a_{s_d}}$.
Within this class, the smallest source order is $m\asymp N^{q_{s_d}}/\eta$;
this sufficient allocation has $n=N-m\sim N$ and source fraction
of order $N^{q_{s_d}-1}/\eta\to0$. Constant positive source fractions
bounded away from one are also rate-optimal. The rate gain over $N$
target observations is of order $N^{a_{s_d}-a_{s_g}}$.
\end{corollary}
Indeed, $0<q_{s_d}<1$ makes the stated minimal-order allocation feasible.
For $n\asymp N$, Theorem 1 gives the upper bound whenever
$M\gtrsim N^{q_{s_d}}$. The discrepancy submodel $g=0$ gives a lower
bound of order $N^{-a_{s_d}}$ for every allocation, even with the source
known, since at most $N$ target observations are available.
Minimality of the source order additionally uses the shared-component
lower bounds: if $M=o(N^{q_{s_d}})$ and $n\asymp N$, the intermediate
risk divided by $N^{-a_{s_d}}$ diverges as
$(N^{q_{s_d}}/M)^{t/\kappa}$, or the low-source rate applies and the ratio
diverges as $N^{a_{s_d}-a_{s_g}}$. Thus smaller source order cannot attain
the optimal order under these fixed-class assumptions. This is an order
statement, not a unique allocation or an optimum of finite-sample constants.
If $\eta$ decreases with $N$, the vanishing fraction for the stated
allocation additionally requires $\eta N^{1-q_{s_d}}\to\infty$.

For \(s_g=1,s_d=2,\beta=0,\kappa=3\), (9) becomes

\[n^{-3/5},\qquad n^{-1/3}M^{-2/3},\qquad n^{-5/7},
\quad\text{with thresholds }n^{2/5},\ n^{4/7}.
\tag{11}\]

The regular and critical cases are proved separately in Supplement A,
Appendix D. If \(\kappa<\beta+1\), target-only and source-known MSE are both of
order \(n^{-1}\). If \(\kappa=\beta+1\), both have order
\(n^{-1}\log n\). A fixed nonzero discrepancy ball prevents a smoother
discrepancy from removing this logarithmic loss. The remaining proofs are in Supplement A, Appendices A--C.

\section{Localized functional
correction}\label{localized-functional-correction}

The following construction exposes the source contribution before
introducing kernel implementation. For integer \(s\ge1\) and
\(0<h\le1/4\), let \(w_{s,h}\) vanish on \([0,h]\), equal \(p_\beta\) on
\([2h,1]\), and on \([h,2h]\) add a polynomial correction of scale
\(h^\beta\). Choose its coefficients so that
\[\int_0^1\{p_\beta(x)-w_{s,h}(x)\}x^j\,dx=0,\qquad 0\le j<s.\] The
moment matrix on \([1,2]\) is positive definite. Thus
\(W_{s,h}f=\int w_{s,h}f\) is well defined, and Taylor's theorem with a
Sobolev remainder yields
\[\|L-W_{s,h}\|_{(H^s)^*}\le C h^{s+\beta+1/2},\qquad
\int_0^1\frac{(p_\beta-w_{s,h})^2}{p_\beta}\le C h^{\beta+1}.\tag{12}\] These
are signed moment weights, not ordinary positive trimming weights.

Write \(Z_j=AY/e_j(X)-(1-A)Y/(1-e_j(X))\). With \(w=w_{s_d,h}\),
consider the analysis-only comparison estimator below, taking $h=h_p$ in
the intermediate regime. It uses the known density and propensities.
\[\widehat\theta_h=\frac1n\sum_T\frac{w(X_i)}{p_\beta(X_i)}Z_{Ti}
+\frac1m\sum_S\frac{p_\beta(X_i)-w(X_i)}{p_\beta(X_i)}Z_{Si}.\tag{13}\] Its
expectation is \(W(g+d)+(L-W)g\), so only \((L-W)d\) remains as bias.
Sobolev embedding bounds the outcome means, and independent sample
variances give \[\sup_{f\in\mathcal F}E_f(\widehat\theta_h-\theta_f)^2
\le C\{\rho^2h^{u_{s_d}}+n^{-1}h^{-t}+M^{-1}h^{\beta+1}\}.\tag{14}\] The
unconditional inverse-weighted observations are not Gaussian. A
conditional Gaussian argument is used only later, after the weights and
design are fixed.

Put \(h_s=n^{-1/D_s}\) and \(h_p=(M/n)^{1/\kappa}\). Equating target and
source variance gives \(h_p\). The conditions \(h_p\ge h_{s_g}\)\ and
\(h_p\le h_{s_d}\)\ are exactly \(M\ge n^{q_{s_g}}\) and
\(M\le n^{q_{s_d}}\). In this intermediate range
\(n h_p^{D_{s_d}}\le1\), so discrepancy bias is no larger than variance.
Below it, target-only recovery gives the better order. Above it,
\(h=h_{s_d}\) reaches the source-known order. The difference
\[q_{s_d}-q_{s_g}=\frac{2\kappa(s_d-s_g)}{D_{s_g}D_{s_d}}>0\] quantifies
the width of the intermediate regime on a logarithmic information scale.

\subsection{Why the bounds match}\label{why-the-bounds-match}

Restrict the experiment to \(b_T=b_S=0\). Its exact likelihood
divergence is \[2\operatorname{KL}(P_{g,d}^{n,m},P_0^{n,m})
=n\int p_\beta e_T(g+d)^2+m\int p_\beta e_Sg^2.\] Smooth boundary bumps
of bandwidth \(h_{s_g}\) with \(d=0\) give the low-source lower bound.
In the intermediate regime use bandwidth \(h_p\) and amplitude
proportional to \((n h_p^{\beta+\kappa+1}+M h_p^{\beta+1})^{-1/2}\). The
shared Sobolev constraint holds because \(h_p\ge h_{s_g}\). Finally, a
discrepancy bump of bandwidth \(h_{s_d}\) with \(g=0\) gives the
high-source lower bound for any source sample size. Each has bounded
divergence and functional separation of the claimed order. Testing
against zero establishes both global and zero-center local lower bounds,
including honest expected-length bounds. No white-noise equivalence is
assumed. Supplement A, Appendices A--D give the full calculations.

Estimating \(Lg\) alone to MSE \(n^{-a_{s_d}}\) would require source
information of order \(n^{a_{s_d}}\). Localization needs only
\(n^{q_{s_d}}\), and \(a_{s_d}-q_{s_d}=(\beta+1)/D_{s_d}>0\). This
comparison does not imply that every two-stage kernel procedure is
slower: some may implicitly estimate the same local residual functional.

\section{Joint RKHS recovery as an
implementation}\label{joint-rkhs-recovery-as-an-implementation}

For a realized design, let \(T_j:H_j\to\mathbb R^{n+m}\) map the \(j\)th
component to observation means, in the order \(b_T,b_S,g,d\). Thus
baseline maps evaluate only within their own sample; \(T_g\) evaluates
multiplied by treatment in both samples; \(T_d\) does so in the target
alone. Let \(R_j=(B,B,R,\rho)_j\), and let \(\ell_j\in H_j\) represent
the component's contribution to \(\theta\). The baseline representers
are zero; the others represent \(L\). Define

\[r_j(v)=\ell_j-T_j^*v,\quad \delta_j(v)=R_j\|r_j(v)\|_{H_j},\quad
\mathcal B(v)=\sum_{j=1}^4\delta_j(v),\quad V(v)=v^\top v.
\tag{15}\]

\begin{proposition}[Quadratic recovery]\label{prop:recovery}

For any \(D\)-measurable \(v\),

\[\sup_{f\in\mathcal F}E_f[(v^\top Y-\theta_f)^2\mid D]
=\mathcal B(v)^2+V(v)=:\mathcal R_D(v).
\tag{16}\]

The quadratic objective and its unique minimizer are

\[Q_D(v)=V(v)+\sum_j\delta_j(v)^2,
\quad \mathsf A=I+\sum_jR_j^2T_jT_j^*,\quad
\mathbf c=\sum_jR_j^2T_j\ell_j,\quad v_Q=\mathsf A^{-1}\mathbf c.
\tag{17}\]

For every \(v\), \(Q_D(v)\le\mathcal R_D(v)\le4Q_D(v)\). In particular,
for every competing weight vector \(w\),

\[\mathcal R_D(v_Q)\le4Q_D(w)\le4\mathcal R_D(w).
\tag{18}\]

\end{proposition}
This is a factor-four comparison within linear estimators for each design. The unconditional rate optimality
follows separately from Theorem 1's lower bound and Supplement A, Appendix F's
expected-\(Q\) bound.

The same estimate is \(L(\widehat g+\widehat d)\), obtained by
minimizing

\[\sum_i\{Y_i-\sum_j(T_j f_j)_i\}^2+\sum_j\|f_j\|_{H_j}^2/R_j^2.
\tag{19}\]

It is therefore existing additive joint KRR/ellipsoidal optimal
recovery. The exact product-ball minimax \emph{linear} estimator instead
minimizes \(V+(\sum \delta_j)^2\). These objectives generally differ. With
average loss over \(n+m\) observations, (19)'s penalties must be divided
by \(n+m\). This rescaling gives the identical objective up to a positive constant.

\paragraph{Inputs to recovery.}
Given the functional $L$, its integral representers, the Sobolev spaces,
radii and noise variance, (17) uses only the realized covariates and
assignments. It uses no propensity values, $\eta$ or $\kappa$.
Pointwise minimization compares it with the analysis weights separately
under each specified design law. However, $L$ and its representers depend
on $p_\beta$, so this is not a guarantee for an estimated, unknown target
density. Conditional honesty needs the correct representers and class
bounds, rather than propensity values. The MSE upper-bound calculations
use independent mean-zero errors with bounded conditional variance;
Gaussianity supplies the exact interval calibration and the stated
likelihood lower bounds.

\subsection{Implementation workflow}\label{implementation-workflow}

The following steps specify the quadratic procedure $v_Q$. The numerical
comparison also evaluates the distinct exact product-ball objective.

\begin{enumerate}
\def\labelenumi{\arabic{enumi}.}
\tightlist
\item
  Fix the target functional, the four Sobolev spaces and radii, and the
  noise variance before using outcomes. Assemble the design operators
  and integral representers in (15).
\item
  Solve (17) for the quadratic estimator $v_Q$. If selecting among
  additional design-only candidates, retain the smallest evaluated
  product-ball risk and include the quadratic candidate. This
  optional selection is distinct from exact optimization over all linear
  weights.
\item
  Evaluate all four dual residual norms using the original class radii.
  Report the estimate, variance, and the sum of the four bias bounds;
  the tuning penalties do not redefine the parameter class.
\item
  Calibrate the interval by (23). Record solver residuals and retain the
  numerical qualification in (22). Selection must depend only on the
  design for the conditional Gaussian argument to apply.
\end{enumerate}

\subsection{Computation for the simulated
model}\label{computation-for-the-simulated-model}

For \(\beta=0\), the representer of \(L\) is the constant function 1 in
each standard \(H^s\). This fact does not hold for general \(\beta\).
For residual \(c\int f-\sum_i v_i f(x_i)\), the exact representer solves

\[-r''+r=c-\sum_i v_i\delta_{x_i},\quad r'(0)=r'(1)=0 \quad(H^1),
\tag{20}\] \[r''''-r''+r=c-\sum_i v_i\delta_{x_i},\quad
r''=0,\ r'-r'''=0\text{ at }0,1 \quad(H^2).
\tag{21}\]

These are natural boundary conditions for the representer, not
restrictions on the unknown functions in (5). Integrating by parts gives
them, as detailed in Supplement A, Appendix G. The implementation propagates the
associated state equations across sorted observations and imposes their
point-source jumps. It uses no fixed-rank truncation. Sorting costs
\(O((n+m)\log(n+m))\); each cached matrix-vector product and its memory
cost are \(O(n+m)\). Conjugate-gradient iteration count is additional
and is not uniformly proved constant.

If \(\xi=\mathsf A\widetilde v-\mathbf c\), then

\[\|\widetilde v-v_Q\|\le\|\xi\|,\qquad
Q_D(\widetilde v)-Q_D(v_Q)=\xi^\top\mathsf A^{-1}\xi\le\|\xi\|^2.
\tag{22}\]

A concrete sufficient absolute stopping rule for the quadratic joint
system is \(\|\xi\|\le\varepsilon_n\sqrt{r(n,M)}\), with bounded
\(\varepsilon_n\) (or \(\varepsilon_n\to0\) for negligible excess
\(Q\)). This preserves the expected-\(Q\) rate by (22). It is a
mathematical sufficient rule, not a guarantee that a fixed-precision
implementation can satisfy it indefinitely.

For inference, bias and variance must be evaluated at the actual
returned weights. Equation (22) alone is not a certificate for exact
floating-point coverage or for the error in the product-ball risk. For
an asymptotic numerical rate claim one needs, for example,
\(E\|\xi\|^2=O(r(n,M))\); a fixed numerical tolerance is not an asymptotic
theorem.

\subsection{Conditional honest confidence
intervals}\label{conditional-honest-confidence-intervals}

Let \(c_\alpha(z)\), for $z\ge0$, solve

\[\Phi(c-z)-\Phi(-c-z)=1-\alpha.\]

For \(V>0\), report

\[C_Q=\left[v_Q^\top Y\ \pm\sqrt{V(v_Q)}
c_\alpha\{\mathcal B(v_Q)/\sqrt{V(v_Q)}\}\right].
\tag{23}\]

For \(V=0\), use half-length \(\mathcal B\). Conditional on \(D\), the
error is Gaussian with variance \(V\) and bias in
\([-\mathcal B,\mathcal B]\). The coverage of a fixed symmetric interval
decreases with the absolute bias, proving conditional honesty. Since
\(c_\alpha(z)\le z+z_{1-\alpha/2}\), expected length is bounded by a
constant times \({E_D\mathcal R_D(v_Q)}^{1/2}\), and hence attains
Theorem 1's order.

Design-only selection of weights or regularization preserves this
argument. Selecting with the same outcomes through cross-validation,
estimating radii from the fitted outcomes, or substituting an
uncalibrated posterior interval does not. Gaussian known-variance errors
and correct deterministic function-class bounds are substantive
conditions. The term robust refers to all discrepancies within the
stated class, not robustness to violating that class. Formula (23) is
the standard bias-aware folded-normal construction.

\subsection{From conditional recovery to the population
rate}\label{from-conditional-recovery-to-the-population-rate}

The pointwise factor-four comparison alone does not establish an
unconditional minimax rate. For the localized moment weights \(v_h\),
bounded Sobolev evaluation norms and the Hilbert-space variance identity
imply
\[E_D Q_D(v_h)\le C\{\rho^2h^{u_{s_d}}+n^{-1}h^{-t}+M^{-1}h^{\beta+1}\}.\]
The baseline residual means
vanish, the two shared-effect residual means cancel, and the discrepancy
residual mean is \(L-W_h\). The centered variances are bounded by the
same inverse-propensity integrals as the noise variance. Pointwise
minimization of \(Q_D\) therefore yields
\[\sup_f E_f(\widehat\theta_Q-\theta_f)^2
\le E_D\mathcal R_D(v_Q)\le4 E_D Q_D(v_h)\lesssim r(n,M).\] Supplement A, Appendix F
supplies the componentwise argument and the target-only comparison for
the low-source regime. The expectation of a conditional supremum is used
as an upper bound, not identified with the unconditional minimax risk.

\section{Numerical illustration}\label{numerical-illustration}

The numerical experiment uses 24 independent random designs per condition. The base
configuration is \(s_g=1,s_d=2,\beta=0,\kappa=3\), \(B=0.5,R=2,\rho=1\),
and \(\eta=1/4\). Noise variance is one. The full standard \(H^1/H^2\)
Green operators are used, without fixed-rank truncation. All displayed
risks evaluate the same product class.

Table 1 reports both numerical minimization of the exact product-ball
linear-risk objective and quadratic recovery (17). The corresponding target-only comparator
minimizes the same objective with source weights absent; it does not
replace the target effect by an \(H^2\) function. Each gain is a ratio
of mean conditional worst-case risks. Equal-budget gains compare
borrowing with \(n+m\) target participants and no source. Monte Carlo
standard errors use paired designs and the delta method.

\begin{table}[ht]
\caption{All nine base cells. Gains are ratios of mean conditional worst risks;
parentheses give paired delta-method MCSE from 24 designs. Both estimators
use their own target-only comparator. Values above one favor borrowing.}
\centering\small
\begin{tabular}{rr r r r r}\toprule
$n$&$m$&\multicolumn{2}{c}{Exact product}&\multicolumn{2}{c}{Quadratic $v_Q$}\\
&&Same target&Equal budget&Same target&Equal budget\\\midrule
128 & 64 & 1.960 (0.082) & 1.514 (0.111) & 1.761 (0.065) & 1.355 (0.100) \\
128 & 256 & 2.500 (0.128) & 1.340 (0.113) & 2.288 (0.105) & 1.213 (0.104) \\
128 & 1024 & 2.823 (0.150) & 0.739 (0.065) & 2.655 (0.132) & 0.682 (0.061) \\
512 & 64 & 1.840 (0.091) & 1.782 (0.096) & 1.667 (0.073) & 1.612 (0.077) \\
512 & 256 & 2.572 (0.152) & 1.953 (0.102) & 2.337 (0.122) & 1.769 (0.083) \\
512 & 1024 & 3.069 (0.202) & 1.420 (0.101) & 2.876 (0.174) & 1.321 (0.093) \\
2048 & 64 & 1.410 (0.044) & 1.403 (0.044) & 1.309 (0.038) & 1.301 (0.039) \\
2048 & 256 & 2.090 (0.104) & 1.925 (0.100) & 1.902 (0.088) & 1.749 (0.086) \\
2048 & 1024 & 2.756 (0.168) & 1.965 (0.117) & 2.569 (0.151) & 1.827 (0.107) \\
\bottomrule\end{tabular}\end{table}

For example, the equal-budget risk reduction is about 49\% at
\((n,m)=(512,256)\), whereas at \((128,1024)\) borrowing has about 35\%
greater risk. These are finite-sample allocation comparisons under equal
participant costs, not uniform dominance. The quantity averaged is
\(E_D\sup_f R(f,D)\), rather than \(\sup_f E_D R(f,D)\). Neither is the
fixed-truth MSE.

For the fixed-truth evaluation, \(b_T(x)=0.12+0.08x\),
\(b_S(x)=-0.1+0.07x\), \(g(x)=1.5|x-0.2|\), and \(d\) is proportional to
\(0.4+0.15x+0.05x^2\), scaled to \(H^2\) norm \(0.8\rho\). At
\(n=512,m=1024\), exact product-ball recovery has mean worst risk
\(0.06570\) and fixed-truth MSE \(0.03850\); quadratic recovery has
\(0.07559\) and \(0.03524\), respectively. Thus the smaller worst risk
does not imply smaller MSE at this truth. Gaussian outcome noise is
integrated analytically; simulation uncertainty comes from random
designs.

The intervals use the declared radii, not estimated radii. When the true
discrepancy radius is one but a smaller bound is supplied, the
sensitivity experiment has minimum fixed-truth coverage about \(0.815\)
and minimum conditional worst-direction coverage about \(0.206\). The
latter direction may change with the design. Correct-class conditional
honesty does not protect against radius misspecification.

The optimization diagnostics give maximum relative
product-risk gap bound \(7.54\times10^{-9}\) and maximum absolute gap
bound \(6.59\times10^{-11}\). These bounds were evaluated in
floating-point arithmetic, not certified interval arithmetic. Designs use nested prefixes and common random numbers. Supplement B retains raw
tables, code, seeds, environment, tuning diagnostics, and failures.
Two-stage comparisons there are specified adaptations, not faithful
replications of Wang et al.'s procedure.

A separate broad-$M$ experiment uses full joint quadratic recovery at
$n=512,2048,8192$, five source levels from approximately
$0.25n^{2/5}$ to $2n^{4/7}$, and 100 designs per condition. For the correctly specified class with $B=0.5$, $R=2$, $\rho=1$
and $\eta=1/4$, reaggregation gives adjacent log-risk slopes from $-0.15$
to $-0.33$, without clear $0,-2/3,0$ transitions. The figure in Supplement A, Appendix H displays these results with MCSE
and nominal thresholds, separately from Table 1's comparisons.
The nominal intermediate band has multiplicative width
$n^{q_{s_d}-q_{s_g}}=n^{6/35}$: approximately $2.9$, $3.7$, and $4.7$
at these three target sizes. This narrow finite-$n$ band permits
constant-factor and transition effects; it does not demonstrate the
asymptotic slope or either breakpoint.

The 24 designs per cell provide exploratory Monte Carlo precision. At $n=512$, $M=64,256$ exceed the nominal upper threshold
$n^{4/7}\approx35$; the full grid includes lower-information cells but
does not validate both asymptotic breakpoints. The numerical
implementation specializes to $\beta=0$ because the integral representer
is constant; the general-$\beta$ statement rests on the proof.

\paragraph{Radius sensitivity.}
The coverage loss to $0.206$ is a worst-direction minimum over designs,
not mean coverage for a single fixed truth. It demonstrates the consequence
of understating the discrepancy class. Adaptation bounds for convex
centrosymmetric classes \citep{ak18,cailow} explain why honest interval
length cannot generally be reduced by estimating a favorable smoothness
bound from the same data. They do not certify an unknown-radius procedure
here. Report $C(\rho)$ across a scientifically justified radius range,
recomputing weights and bias bounds at each value, and report where any
sign conclusion changes. This is sensitivity analysis rather than
outcome-based selection of the shortest interval. Supplement B reports
the existing radius grid, mean lengths and coverage diagnostics.

\section{Discussion}\label{discussion}

Source information recovers the local residual functional left uncertain
by target observations. Its two thresholds separate the onset of rate
improvement from the source-known rate. Fixed-budget allocation follows
at the rate level. The equal-budget loss at $(128,1024)$ is a finite-sample
risk comparison and does not establish asymptotic over-allocation.
A fixed source fraction below one can be rate-optimal; if $n/N\to0$,
the discrepancy lower bound prevents the optimal $N^{-a_{s_d}}$ order.
Asymptotic rates alone do not determine the finite-sample optimum.

The common covariate density and uniformly informative source near the
boundary are substantive assumptions: they give the source variance
$M^{-1}h^{\beta+1}$. A different source density or vanishing source overlap
changes that integral and requires new upper and lower bounds. Smoothness
ordering is equally substantive. With fixed positive radii and
$s_d\le s_g$, target-only and source-known rates are both governed by
$s_d$, so there is no rate-order improvement of this type. A possible
scientific rationale for $s_d>s_g$ is a shared response with a localized
kink and a domain change that shifts it by a smooth trend on the same
covariate scale; this remains an assumption to justify in applications.

With zero target treatment on $[0,h_0)$, Supplement A, Appendix E instead gives
\[
\mathfrak R^{\rm hole}_{n,m}\asymp n^{-1}+\rho^2h_0^{u_{s_d}}
+\min\{R^2h_0^{u_{s_g}},h_0^{\beta+1}/M\}.\tag{24}
\]
For fixed $h_0>0$ the discrepancy leaves identification uncertainty even
with unlimited source information. The zero-center identified interval
has half-width $B_{s_d}(\rho,h_0)$, where
$B_s(r,h_0)\asymp r h_0^{s+\beta+1/2}$, and is centered at zero.
This differs from the point-identified weak-overlap setting of Theorem 1.

\section*{Competing interests}
The author declares no competing interests.

\section*{Declaration of generative AI use}
During the preparation of this work, the author used ChatGPT and Codex
to assist with mathematical derivations and their review, code development,
numerical checks, and manuscript editing. The author reviewed and edited
the content as necessary and takes full responsibility for the content
of the publication.

\begin{supplement}\stitle{Supplement A: Complete proofs}\sdescription{Moment bounds, testing lower bounds, regular and critical regimes, zero-overlap identification, conditional recovery, and Green operators.}\end{supplement}
\begin{supplement}\stitle{Supplement B: Reproducibility archive}\sdescription{Code, raw results, seeds, environment, and reproduction instructions are bundled as Supplement B in the accompanying source package (anc/supplement\_B\_v9.zip). The broad-source-information figure is also included in Supplement A.}\end{supplement}

\clearpage\section*{Supplement A: Complete proofs}
\noindent Equation numbers (1)--(24) refer to the main manuscript. Letter-prefixed equation numbers below are internal to this supplement. Equations (E1)--(E3) below concern the supplementary hole theorem.
\section*{Appendix A Moment weights and their
bounds}\label{appendix-a-moment-weights-and-their-bounds}

For \(0<h\le1/4\) and integer \(s\ge1\), define \(w_{s,h}=0\) on
\([0,h]\), \(w_{s,h}=p_\beta\) on \([2h,1]\), and on \([h,2h]\) set

\[w_{s,h}(x)=p_\beta(x)+h^\beta\sum_{k=0}^{s-1}c_k(x/h)^k.\]

Choose the coefficients by

\[\sum_{k=0}^{s-1}c_k\int_1^2z^{j+k}\,dz
=\frac{\beta+1}{\beta+j+1},\quad j=0,\ldots,s-1.
\tag{A1}\]

The coefficient matrix is a polynomial Gram matrix on \([1,2]\), hence
positive definite: its quadratic form at a nonzero coefficient vector
equals the strictly positive integral of the corresponding polynomial
squared. A unique solution exists and is independent of \(h\).
Substitution \(x=hz\) proves

\[\int_0^1(p_\beta-w_{s,h})x^j\,dx=0,\quad j<s.
\tag{A2}\]

For \(f\in H^s\), its traces through order \(s-1\) at zero exist.
Subtract its Taylor polynomial \(P\) of order \(s-1\). The integral
remainder and Cauchy--Schwarz give

\[|f(x)-P(x)|\le\frac{x^{s-1/2}}{(s-1)!\sqrt{2s-1}}\|f^{(s)}\|_{L^2(0,x)}.\]

The signed residual measure is supported on \([0,2h]\), has total
variation \(O(h^{\beta+1})\), and annihilates \(P\). Therefore, writing
\(\mathcal E_{s,h}=L-W_{s,h}\) with \(W_{s,h}f=\int w_{s,h}f\),

\[\|\mathcal E_{s,h}\|_{(H^s)^*}\le C h^{s+\beta+1/2}.
\tag{A3}\]

Moreover,

\[\int\frac{(p_\beta-w_{s,h})^2}{p_\beta}\,dx\le Ch^{\beta+1},
\quad
\int_h^1\frac{w_{s,h}^2}{p_\beta}
\left(\frac1{e_T}+\frac1{1-e_T}\right)dx
\le C\left(1+\int_h^1x^{\beta-\kappa}dx\right).
\tag{A4}\]

On \([0,h]\), the first integrand is \(p_\beta\). On \([h,2h]\), the
substitution \(x=hz\), bounded polynomial coefficients, and
\(z\in[1,2]\) give both bounds. On \([2h,1]\), use \(w=p_\beta\),
\(e_T=x^\kappa/2\), and \(1-e_T\ge1/2\). Likewise, since \(e_S\ge\eta\)
and \(1-e_S\ge1/3\), the source integral with reciprocal propensities is
at most \(Ch^{\beta+1}/\eta\). These arguments also apply to a weight
with more moments when (A3) is required for a lower integer smoothness
order.

\section*{Appendix B Upper bounds and the
thresholds}\label{appendix-b-upper-bounds-and-the-thresholds}

Let \(Z_j=AY/e_j(X)-(1-A)Y/(1-e_j(X))\). At points of zero propensity a
product with a zero integration weight is defined as zero; the
weak-overlap zero endpoint has probability zero. Direct conditioning
gives \(E[Z_T\mid X]=g+d\), \(E[Z_S\mid X]=g\). Sobolev embedding on the
compact interval bounds all outcome means uniformly over (5), so

\[E[Z_j^2\mid X]\le C\{e_j(X)^{-1}+(1-e_j(X))^{-1}\}.
\tag{B1}\]

For \(w=w_{s_d,h}\), use

\[\widehat\theta_{J,h}=\frac1n\sum_T\frac{w(X_i)}{p_\beta(X_i)}Z_{Ti}
+\frac1m\sum_S\frac{p_\beta(X_i)-w(X_i)}{p_\beta(X_i)}Z_{Si}.
\tag{B2}\]

Its expectation is \(W(g+d)+(L-W)g=\theta-\mathcal E_{s_d,h}d\). Independence and (B1),
using second moments as variance upper bounds, imply

\[\sup_f E_f(\widehat\theta_{J,h}-\theta)^2
\le C\{\rho^2h^{u_{s_d}}+n^{-1}h^{-t}+M^{-1}h^{\beta+1}\},\quad t>0.
\tag{B3}\]

Known upper bounds on absolute bias and variance also yield an
unconditional honest interval with half-length
\(B_d(h)+\sqrt{V_J(h)/\alpha}\) by Chebyshev's inequality. The
unconditional pseudo-outcome error is not assumed Gaussian.

Target-only uses the first sum in (B2) with \(w_{s_g,h}\). Its bias is
\(\mathcal E_{s_g,h}(g+d)\), bounded by a constant times \(h^{s_g+\beta+1/2}\) because
\(H^{s_d}\subset H^{s_g}\). Setting \(h_g=n^{-1/D_{s_g}}\) gives risk
\(O(n^{-a_{s_g}})\). A source-known oracle adds \(\mathcal E_{s_d,h}g\) to the target sum
with \(w_{s_d,h}\), and \(h_d=n^{-1/D_{s_d}}\) gives
\(O(n^{-a_{s_d}})\).

For finite source, let \(h_p=(M/n)^{1/\kappa}\). The equality
\(n^{-1}h_p^{-t}=M^{-1}h_p^{\beta+1}\) follows from
\(t+\beta+1=\kappa\). Its common value is
\(n^{-(\beta+1)/\kappa}M^{-t/\kappa}\). The comparisons \(h_p\ge h_g\)
and \(h_p\le h_d\) are respectively equivalent to \(M\ge n^{q_{s_g}}\)
and \(M\le n^{q_{s_d}}\). In this range,

\[\frac{h_p^{u_{s_d}}}{n^{-1}h_p^{-t}}=n h_p^{D_{s_d}}\le1.\]

Above the upper threshold choose \(h_d\); its source variance is at most
its target variance because \(M\ge n h_d^\kappa=n^{q_{s_d}}\). Below the
lower threshold use target-only. Methods and bandwidths in this proof
are chosen from known parameters, not outcomes. Finite initial sample
sizes can be covered by constant enlargement and truncation of
bandwidths to \(1/4\). This proves the MSE and interval upper rates in
Theorem 1.

\section*{Appendix C Lower bounds and local
meaning}\label{appendix-c-lower-bounds-and-local-meaning}

Restrict to \(b_T=b_S=0\). Conditional Gaussian likelihoods and the
common design law give the exact identity

\[2\mathrm{KL}(P_{g,d}^{n,m},P_0^{n,m})
=n\int p_\beta e_T(g+d)^2+m\int p_\beta e_Sg^2.
\tag{C1}\]

This is KL for the full random-design experiment, not an assertion of
white-noise equivalence. It holds because KL conditional on each design
point is one half the squared mean difference; expectation over that
point and additivity over independent observations give (C1).

Fix a nonnegative, nonzero \(\psi\in C_c^\infty(0,1)\). For
\(f_h=A\psi(x/h)\),

\[\|f_h\|_{H^s}^2\le C A^2h^{1-2s},\quad
Lf_h=c_\psi A h^{\beta+1},\quad
\int x^\nu f_h^2=C_{\psi,\nu}A^2h^{\nu+1}.
\tag{C2}\]

All constants are positive where required and independent of
\(h\le1/4\). Choose a fixed small amplitude multiplier so that the norm
bounds and KL bound \(c_0\) hold.

First set \(g=0\), \(d=c h_d^{s_d-1/2}\psi(x/h_d)\). By (C1), KL is
bounded by \(Cc^2 n h_d^{D_{s_d}}=Cc^2\), while the squared functional
separation has order \(n^{-a_{s_d}}\). This proves the discrepancy lower
bound for every \(M\), including the source-known experiment.

Next set \(d=0\), \(g=c h_g^{s_g-1/2}\psi(x/h_g)\). Target KL is bounded
and source KL is at most \(Cc^2M h_g^{2s_g+\beta}\). This is bounded for
\(M\le n^{q_{s_g}}\), and the squared separation has order
\(n^{-a_{s_g}}\). For target-only the same direction proves its lower
bound without a source KL restriction.

In the intermediate range take \(h=h_p\), \(d=0\), and

\[A=\frac{c}{\sqrt{n h^{\beta+\kappa+1}+M h^{\beta+1}}}.
\tag{C3}\]

Equation (C1) is bounded by \(Cc^2\). As \(h\ge h_g\),

\[\|g\|_{H^{s_g}}^2\le\frac{Cc^2}{n h^{D_{s_g}}}\le Cc^2.\]

Thus this alternative belongs to the fixed ball. Its squared separation
is

\[(Lg)^2\asymp\frac{h^{2\beta+2}}{n h^{\beta+\kappa+1}+M h^{\beta+1}}
\asymp n^{-1}h^{-t}=n^{-(\beta+1)/\kappa}M^{-t/\kappa}.
\tag{C4}\]

To state the testing step explicitly, if two parameter values have
functional separation \(\Delta\), an estimator determines the
nearest-parameter test. A wrong decision entails squared error at least
\(\Delta^2/4\). The sum of testing errors is at least
\(1-\mathrm{TV}(P_0,P_1)\), hence the maximum MSE is at least
\(\Delta^2(1-\mathrm{TV})/8\). Pinsker's inequality and sufficiently
small \(c\) keep TV below a fixed constant less than one. The
alternatives above and zero all belong to \(\mathcal U_{c_0}(0)\),
proving both global and local lower bounds. Taking the maximum of the
discrepancy and shared lower bounds gives (9); no simultaneous
least-favorable direction is needed.

For any globally honest interval, under \(P_0\) the probability that
both parameter values lie in the interval is at least
\(1-2\alpha-\mathrm{TV}(P_0,P_1)\). Its expected length at zero is
therefore at least this positive constant times \(\Delta\). Choose \(c\)
small enough depending on \(\alpha,c_0\). This proves the local and
global interval lower bounds. The global upper bound also bounds risk on
the local set and length at zero, completing Theorem 1's quantifiers.

\section*{Appendix D Regular and critical
overlap}\label{appendix-d-regular-and-critical-overlap}

If \(\kappa<\beta+1\), untruncated IPW has uniformly bounded second
moment up to a constant because \(\int_0^1x^{\beta-\kappa}dx<\infty\).
It gives risk \(O(n^{-1})\); a constant discrepancy of size
\(c/\sqrt n\), with \(g=0\), has bounded KL and squared functional
separation \(c^2/n\).

If \(\kappa=\beta+1\), (A4) is \(O(1+\log(1/h))\). Choosing
\(h=n^{-1/u_s}\) for the relevant smoothness gives \(O(\log n/n)\). For
the lower bound choose \(s=s_d\), \(g=0\),

\[d_h(x)=a(x+h)^{-\kappa},\quad h=n^{-1/(2s+2\beta+1)},\quad
a=c\{n\log(1/h)\}^{-1/2}.\]

The largest Sobolev derivative integral is
\(O(a^2h^{1-2s-2\kappa})=O(c^2/\log(1/h))\), so the fixed ball condition
holds. Target KL is \(O(na^2\log(1/h))=O(c^2)\); source KL is zero.
Since \(p_\beta(x)(x+h)^{-\kappa}\) has logarithmic integral, the
squared functional separation is of order \(\log n/n\). The same testing
argument gives both MSE and interval lower bounds. Target-only uses the
analogous construction with \(s=s_g\). This proves the statements
following Theorem 1.

\section*{Appendix E Hole risk and
identification}\label{appendix-e-hole-risk-and-identification}

\begin{theorem}[Zero-overlap hole]

Replace (2)'s target propensity by zero on \([0,h_0)\) and \(1/2\) on
\([h_0,1]\), where \(0<h_0\le1/4\). Keep the other assumptions and
\(s_d>s_g\). Fix the positive radii \(B,R,\rho\) and the smoothness and
density parameters. There exist constants \(c,C,n_0\), independent of
\(h_0,m,\eta\), such that for every \(n\ge n_0\), \(0<h_0\le1/4\), and
\(M=m\eta\ge1\), the risk is bounded above and below by constant
multiples of

\[\mathfrak R_{n,m}^{\mathrm{hole}}\asymp
n^{-1}+\rho^2h_0^{u_{s_d}}+
\min\{R^2h_0^{u_{s_g}},h_0^{\beta+1}/M\}.
\tag{E1}\]

The corresponding minimax honest expected interval length is the square
root of (E1). The bounds are uniform in \(0<h_0\le1/4\) for fixed
positive radii. They are also uniform when all radii range in prescribed
compact subsets of \((0,\infty)\), with constants and \(n_0\) depending
on those subsets. No uniform claim over radii approaching zero is made.
The fixed-hole case is the identification interpretation.

At the infinite-data zero center, define

\[B_s(r,h_0)=\sup\{|Lf|:\|f\|_{H^s}\le r,\ f=0\ \text{a.e. on }[h_0,1]\}.
\tag{E2}\]

With both domains observed, the identified set is exactly

\[[-B_{s_d}(\rho,h_0), B_{s_d}(\rho,h_0)],\qquad
B_s(r,h_0)\asymp r h_0^{s+\beta+1/2}.
\tag{E3}\]

The full identification width is\ twice \(B_{s_d}\). The
target-only half-width is of order
\(Rh_0^{s_g+\beta+1/2}+\rho h_0^{s_d+\beta+1/2}\). This last expression
is an order statement, not an exact sum: shared and discrepancy
components can cancel outside the hole. Section 3's weak-overlap model
remains point identified; its slow rates are not identification widths.
\end{theorem}
The remainder of this appendix proves these assertions.

For the hole propensity, take \(h=h_0\) in (B2). All nonzero target
weights are supported where \(e_T=1/2\). Equations (A3)--(A4), now
without the diverging inverse propensity, give risk at most
\(C\{n^{-1}+\rho^2h_0^{u_{s_d}}+h_0^{\beta+1}/M\}\). For target-only use
\(w_{s_d,h_0}\), which also annihilates polynomials through order
\(s_g-1\). Bounding its residual separately on \(g\) and \(d\), and
using \((a+b)^2\le2a^2+2b^2\), gives
\(C\{n^{-1}+R^2h_0^{u_{s_g}}+\rho^2h_0^{u_{s_d}}\}\). Taking the smaller
bound proves the upper part of (E1).

The moment constants above are independent of \(h_0\le1/4\). On any
prescribed compact set of positive radii, the upper constants are
bounded and a single \(n_0\) ensures the constant discrepancy below lies
inside the ball. If radii instead vanish with \(n\), this argument no
longer supplies a uniform \(n^{-1}\) term: for example \(R=\rho=n^{-1}\)
bounds the squared target functional by \(O(n^{-2})\), contradicting a
uniform \(c/n\) lower bound.

Three two-point constructions give its lower bound. A constant
discrepancy of size \(c/\sqrt n\) gives \(n^{-1}\), since the treated
mass outside a hole of size at most \(1/4\) is bounded below. A
discrepancy bump \(c\rho h_0^{s_d-1/2}\psi(x/h_0)\) has zero observation
KL and squared functional separation of order \(\rho^2h_0^{u_{s_d}}\).
Finally a shared bump in the hole of amplitude

\[c\min\{R h_0^{s_g-1/2},(M h_0^{\beta+1})^{-1/2}\}\]

has zero target KL, bounded source KL, and squared separation of order
\[\min\{R^2h_0^{u_{s_g}},h_0^{\beta+1}/M\}.\] The maximum of three
nonnegative terms is at least one third their sum. The MSE and interval
testing arguments of Appendix C complete (E1).

At the infinite-data zero center, source controls and treated outcomes
identify \(b_S=g=0\) throughout the interval, while target controls
identify \(b_T=0\). The treated target observations force \(d=0\) on
\([h_0,1]\). The remaining feasible set for \(d\) is the ball
intersected with this closed linear subspace. It is weakly compact,
convex and symmetric; \(L\) is continuous, so its image is the closed
interval in (E3). Functions in the subspace have zero traces at \(h_0\)
through derivative order \(s-1\). Taylor expansion backwards from
\(h_0\) bounds their integral by \(Crh_0^{s+\beta+1/2}\); (C2) gives the
matching lower bound. This proves the order in (E3).

Without source observations, feasibility only forces \(g+d=0\) outside
the hole. The preceding moment bound gives the asserted sum upper order;
choosing independent compactly supported bumps with the same sign gives
the matching sum lower order. It does not yield exact equality of the
separate support functions. The indistinguishable endpoints in (E3)
further imply \(E|C|\ge2B_{s_d}(\rho,h_0)(1-2\alpha)\) for any pointwise
honest interval under their common law. This is not a claim that the
entire identified set is simultaneously covered with probability
\(1-\alpha\).

\section*{Appendix F Conditional recovery and its unconditional
rate}\label{appendix-f-conditional-recovery-and-its-unconditional-rate}

For fixed \(D,v\), bias equals \(-\sum_j\langle r_j(v),f_j\rangle\).
Cauchy--Schwarz bounds its absolute value by \(\sum_jR_j\|r_j\|\), and
taking each nonzero \(f_j=R_jr_j/\|r_j\|\) makes all signs agree. Zero
residual components may be set to zero. This proves the exact supremum
(16). With \(\delta_j\ge0\),

\[\sum_j\delta_j^2\le(\sum_j\delta_j)^2\le4\sum_j\delta_j^2,\]

which proves (18), retaining the nonnegative variance term. Expanding
\(Q\) gives \(Q(v)=\mathrm{const}-2\mathbf c^\top v+v^\top\mathsf Av\);
\(\mathsf A\succeq I\) proves uniqueness and (17). The normal equations
for (19) give \(\widehat f_j=R_j^2T_j^*\mathsf A^{-1}Y\), hence the
target functional equals \(\mathbf c^\top\mathsf A^{-1}Y=v_Q^\top Y\). This
proves the KRR identity and (22) by completing the square.

Here is the explicit bridge from random-design moment estimation
to expected conditional worst risk. Put

\[q_j(X,A)=A/e_j(X)-(1-A)/(1-e_j(X)),
\quad a_T=w/p_\beta,\quad a_S=(p_\beta-w)/p_\beta.\]

The comparison weights are \(v_i=a_T(X_i)q_T(X_i,A_i)/n\) in target and
\(a_Sq_S/m\) in source. Conditional on \(X\), \(E q_j=0\),
\(E(Aq_j)=1\),

\[E q_j^2=e_j^{-1}+(1-e_j)^{-1},\qquad E(Aq_j)^2=e_j^{-1}.
\tag{F1}\]

For independent identically distributed Hilbert-valued variables \(U_i\)
with finite second moment,

\[E\left\|\ell-\frac1k\sum_iU_i\right\|^2
=\|\ell-EU_1\|^2+\frac1k E\|U_1-EU_1\|^2.
\tag{F2}\]

This follows by expanding around the mean; cross terms vanish by
independence and centering. All Sobolev evaluation kernels here have
bounded diagonal, say \(k_s(x,x)\le C_s\), from the bounded evaluation
norm on a compact interval.

Apply (F2) to baseline atoms \(a_jq_j k_{1,X}\) and treated-effect atoms
\(a_jAq_j k_{s,X}\), separately in the independent samples. Baseline
expected residuals are zero by \(Eq_j=0\). The shared expected residual
is \(\ell_g-W_g-(L-W)_g=0\). The discrepancy expected residual is the
representer of \(\mathcal E_{s_d,h}\). Each centered second moment is at most its
uncentered counterpart, bounded by a kernel diagonal constant times the
integrals in (A4) using (F1). The variance term \(E\sum_i v_i^2\) has
exactly the same integral upper bound. Consequently

\[E_DQ_D(v_{J,h})\le C\{\rho^2h^{u_{s_d}}+n^{-1}h^{-t}+M^{-1}h^{\beta+1}\}.
\tag{F3}\]

For target-only weights, the expected residual for \(g\) and \(d\) is
\(\mathcal E_{s_g,h}\) in their respective spaces, and the same calculation
yields \(E_DQ_D(v_T)\le C\{h^{u_{s_g}}+n^{-1}h^{-t}\}\). Pointwise
minimization of \(Q\), followed by (18), (F3), and Appendix B's choices
now proves

\[\sup_f E_f(\widehat\theta_Q-\theta_f)^2
\le E_D\mathcal R_D(v_Q)\lesssim r(n,M).
\tag{F4}\]

For exact design-only tuning over a candidate family containing the
quadratic joint solution, minimizing the same product-ball risk
yields \(\mathcal R_D(v_{\mathrm{sel}})\le\mathcal R_D(v_Q)\). Thus the
rate and interval-length conclusions also hold for that ideal selection.
Numerical selection requires control of optimization and evaluation
errors; finite-search success is not a global optimality certificate. No
analogous rate claim for the source-first two-stage family follows from
this argument.

No supremum and expectation are interchanged. Applying Cauchy--Schwarz
to \(E_D\mathcal B\) and \(E_D\sqrt V\) proves the interval upper order
in the main inference subsection. The identical argument, with target
inverse propensity bounded on the support of the weights, applies to the
hole model. Proposition 1 and its rate connection are proved.

\section*{Appendix G Green operator and coverage
derivations}\label{appendix-g-green-operator-and-coverage-derivations}

For \(H^1\), integration by parts in \(\int(rf+r'f')\) gives
\(\int(r-r'')f+[r'f]_0^1\). Since test functions have arbitrary endpoint
traces, the boundary terms vanish exactly when \(r'=0\) at both
endpoints. A point mass \(-v_i\delta_{x_i}\) then implies the jump
\(r'(x_i+)-r'(x_i-)=v_i\), with \(r\) continuous.

For \(H^2\), twice integrating the second-derivative term gives

\[\langle r,f\rangle_{H^2}=
\int(r-r''+r'''')f+[r''f'+(r'-r''')f]_0^1.
\tag{G1}\]

Independent endpoint traces of \(f,f'\) yield \(r''=0\), \(r'-r'''=0\).
Interior continuity of\ \(r,r',r''\) and jump
\(r'''(x_i+)-r'''(x_i-)=-v_i\) yield (21). For \(s=1\) the homogeneous
state matrix is
\(\bigl(\begin{smallmatrix}0&1\\1&0\end{smallmatrix}\bigr)\); for
\(s=2\) its last row is \((-1,0,1,0)\), with superdiagonal ones. The
constant particular solution is \(r=c\). These matrices, jumps, and
boundary subspaces match the implementation. Finally

\[\|r\|_{H^s}^2=c\int r-\sum_i v_i r(x_i)
\tag{G2}\]

gives the dual norm. The \(H^1\) kernel independently equals
\[\cosh(\min(x,y))\cosh(1-\max(x,y))/\sinh(1).\] Boundary conditions at
observations exactly equal to 0 or 1 require the corresponding weak-form
interpretation; random continuous designs have no such atoms almost
surely.

For coverage, define \(G(c,z)=\Phi(c-z)-\Phi(-c-z)\). It is even in
\(z\), and for \(c,z\ge0\), \(\partial_zG=\phi(c+z)-\phi(c-z)\le0\).
Hence calibration at \(z=\mathcal B/\sqrt V\) covers every allowed mean
shift. At \(c=z+z_{1-\alpha/2}\), the coverage is at least \(1-\alpha\),
proving the stated additive bound. Every weight choice in this argument
is fixed after conditioning on design. Outcome-dependent tuning
invalidates this fixed-weight Gaussian calculation without further
argument.

\clearpage
\section*{Appendix H Broad-source-information numerical results}
\begin{figure}[ht]
\centering\includegraphics[width=\linewidth]{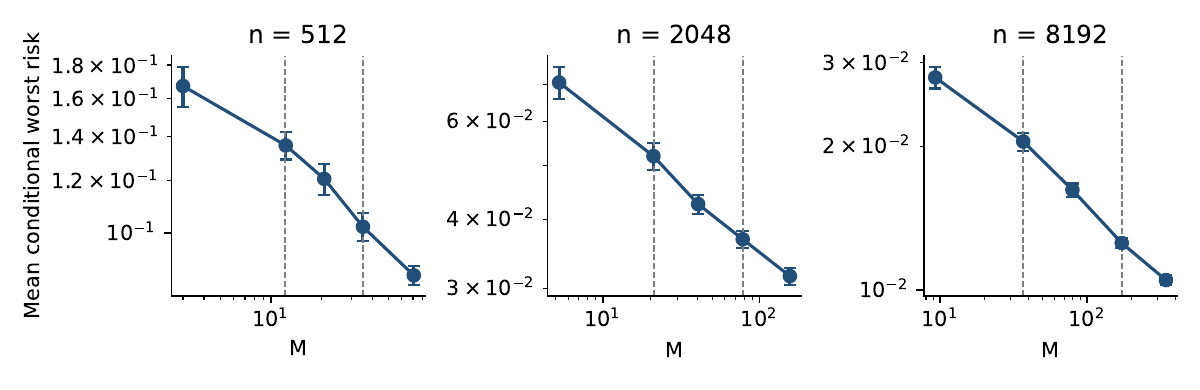}
\caption{Full joint quadratic recovery with $s_g=1$, $s_d=2$, $\beta=0$,
$\kappa=3$, $\eta=1/4$, $B=0.5$, $R=2$, and $\rho=1$.
Each point averages conditional worst risk over 100 random designs;
error bars are $\pm1.96$ Monte Carlo standard errors. Dashed lines mark
$n^{2/5}$ and $n^{4/7}$. The existing numerical results do not resolve
sharp breakpoints. The curves are not minimax risks or fixed-truth MSEs.}
\label{fig:broadM}
\end{figure}
The nominal intermediate widths are $n^{6/35}\approx2.9,3.7,4.7$ for
$n=512,2048,8192$. Adjacent empirical log-risk slopes range from
$-0.15$ to $-0.33$; these are descriptive secants, not estimates of an
asymptotic exponent. Supplement B contains the underlying raw results,
aggregates, seeds, code, and numerical qualifications.
\end{document}